\documentclass[]{spie}  

\usepackage{standalone}
\usepackage{comment}
\usepackage{cite}
\usepackage{algorithmic}
\usepackage{amsmath,amssymb,amsfonts}
\usepackage{graphicx}
\graphicspath{ {./images/} {./figures/} }
\usepackage[colorlinks=true, allcolors=blue]{hyperref}
\usepackage{subcaption}
\usepackage{pgf}
\usepackage{pgfplots}
\usepgfplotslibrary{external}
\usepackage{textcomp}
\usepackage{xcolor}
\usepackage{datetime}
\usepackage{placeins}
\usepackage{tikz}
\usepackage{tikz-3dplot}
\usetikzlibrary{calc,arrows.meta,positioning,backgrounds, matrix}
\usepackage{setspace}
\usepackage{adjustbox}

\title{Improving spatial domain based image formation through compressed sensing.}

\author[a, b]{Gene Stoltz}
\author[b]{Andr\'{e} Leon Nel}
\affil[a]{Council for Scientific and Industrial Research of South Africa}
\affil[b]{University of Johannesburg, South Africa}

\authorinfo{Further author information: (Send correspondence to G.Stoltz)\\G.Stoltz: E-mail: ggsgene@gmail.com}

\begin{document}
\maketitle

\begin{abstract}
In this paper, we improve image reconstruction in a single-pixel scanning system by selecting an detector optimal field of view. Image reconstruction is based on compressed sensing and image quality is compared to interpolated staring arrays. The image quality comparisons use a ``dead leaves" data set, Bayesian estimation and the Peak-Signal-to-Noise Ratio (PSNR) measure.
    
Compressed sensing is explored as an interpolation algorithm and shows with high probability an improved performance compared to Lanczos interpolation. Furthermore, multi-level sampling in a single-pixel scanning system is simulated by dynamically altering the detector field of view. It was shown that multi-level sampling improves the distribution of the Peak-Signal-to-Noise Ratio.

We further explore the expected sampling level distributions and PSNR distributions for multi-level sampling. The PSNR distribution indicates that there is a small set of levels which will improve image quality over interpolated staring arrays. We further conclude that multi-level sampling will outperform single-level uniform random sampling on average.
\end{abstract}


\section{Introduction}
Since the development of compressed sensing, the image formation process within the optical domain can be divided into two categories; spatial domain sampling and global domain sampling. Global domain sampling can be defined as the sampling of the total output image field of view by transforming the wavefront into a basis such as single pixel imaging using a Digital Mirror Device~\cite{Duarte2008} or multiple scatter imaging using scattering media~\cite{Liutkus2014}. Spatial domain sampling, the traditional method, is achieved by using the irradiance measurement of an optical detector with a specific field of view. Multiple of these measurements can be used to reconstruct an output image much larger than the detector field of view with the measurements linearly correlated to the output image. The two most notable methods are the use of a staring array and the scanning of a single pixel such as in a Rosette Imager~\cite{stoltz2020performance} using a Rosette scanner or a scanning optical microscopy using an optical probe~\cite{Seibel1997}.

In Rosette imaging, compressed sensing has been used to transform non-grid based spatial domain samples into grid-based samples, simulating a staring array~\cite{Uzeler2016}. Using the Rosette Imager, it has been shown that non-grid based sampling can match the performance of a traditional staring array (grid-based sampling)~\cite{stoltz2020performance}. These results are not unexpected as compressed sensing require signals to be sparse in the sampling domain and natural scenes are not sparse in the spatial domain.

In recent years, compressed sensing principles have been expanded from incoherence, sparsity, and random sampling to more general principles; asymptotic sparsity, asymptotic incoherence, and multilevel sampling~\cite{Adcock2017}, with multi-level uniform sampling further extended to structured sampling~\cite{Li2016}. With these expanded principles we investigate compressed sensing firstly with regards to structured sampling and secondly with regards to multi-level sampling. Structured sampling is explored through grid-based sampling in the spatial domain and can be implemented by equivalent hardware such as Single Fiber Endoscopy(SFE)~\cite{Seibel2008}. Secondly, multi-level sampling in the spatial domain is explored by assuming the change in detector field of view simulates different sampling levels which can be implemented in systems such as a Rosette Imager.

In Section~\ref{sec:single}, the compressed sensing setup for spatial domain sampling is discussed. It is shown how compressed sensing can be used for interpolation, how an optimal probe size can be selected for single pixel scanning systems and a multi-level sampling approach is developed. In Section~\ref{sec:method}, the tools to evaluate image quality and compare different algorithms are discussed. In Section~\ref{sec:results} the comparative results for interpolation and multi-level sampling are presented. Finally,  in Section~\ref{sec:discuss}, the results are discussed and possible future work is presented.

\section{Single Detector Spatial Domain Image Formation}
\label{sec:single}
The image formation process in the spatial domain uses the irradiance measurements of single detectors and related directional positions. A staring array can be simulated by using a square detector and moving the detector in a grid-like pattern until all positions have been covered correlating each detector position to a pixel in the output image. The single detector scanning system does not always have a one-to-one relation to the output image, such as probe overlap when using circular probes, and therefore regularisation methods are required for image formation.

Compressed sensing provides a regularisation method to reconstruct an image from overlapping probes, different probe shapes, and unstructured sampling positions; and is elaborated on in Section~\ref{sec:cs_setup}. The capability of a single detector to simulate a staring array creates the opportunity for compressed sensing to function as an interpolation algorithm. Compressed sensing as an interpolation algorithm as well as the use of an optimal square or circular probe size in grid-based sampling is explored in Section~\ref{sec:optimal}.

Compressed sensing theory suggests multi-level sampling to improve signal reconstruction. Levels within the spatial domain is ill-defined compared to levels in the Wavelet Transform. In Section~\ref{sec:multi} we define levels equivalent to different probe sizes and develop an approach to multi-level uniform random sampling.

\subsection{Compressed Sensing}
\label{sec:cs_setup}
The compressed sensing framework requires a measurement matrix and a regularisation algorithm within a sparsity basis. In this paper the measurement matrix is formed through spatial domain sampling~\cite{Uzeler2016}. A sample within the measurement matrix is constructed by discretizing a probe onto the output image at a predetermined position. The probe simulates the irradiance measurement of a detector taking the sum of the pixels covered in the output image.

The regularisation algorithm is chosen to support real-time reconstruction from structured sampling and thus the Fourier Domain Regularisation Inversion~(FDRI) algorithm is an adequate choice~\cite{stoltz2020performance}. The FDRI algorithm only approximates an $l1$-optimisation algorithm but does match the performance of Nesterov's algorithm (NESTA) and Basis Pursuit Denoising~(BPDN)\cite{Czajkowski2018}.

It is noted that in the the paper an output image size of 64x64 will be used with 19.1\% sampling, equating to 784 samples. The 784~samples allow for a 28x28 sampling grid, equating to a 28x28 staring array with appropriate square probe, as explained in Section~\ref{sec:optimal}.

\subsection{Grid-based Sampling}
\label{sec:optimal}
The grid-based sampling method simulate detector positions similar to elements in a staring array where various probe forms and sizes can be used to sample at these positions. Mapping a sampling grid of 28x28 onto an output image of 64x64 pixels using square probes of 2.28x2.28 pixels placed on the sampling grid to simulate a 28x28 staring array. The measurement matrix can be created by using each square probe as an entry. This measurement matrix can then be used to generate the inversion matrix using the FDRI algorithm. The compressed sensing framework effectively becomes an interpolation algorithm as developed in the this section. The interpolation algorithm is called the \textit{CS-FPA}.

\begin{figure}[h!]
    \centering
    \begin{subfigure}{0.5\textwidth}
        \includegraphics[width=\columnwidth,interpolate=false]{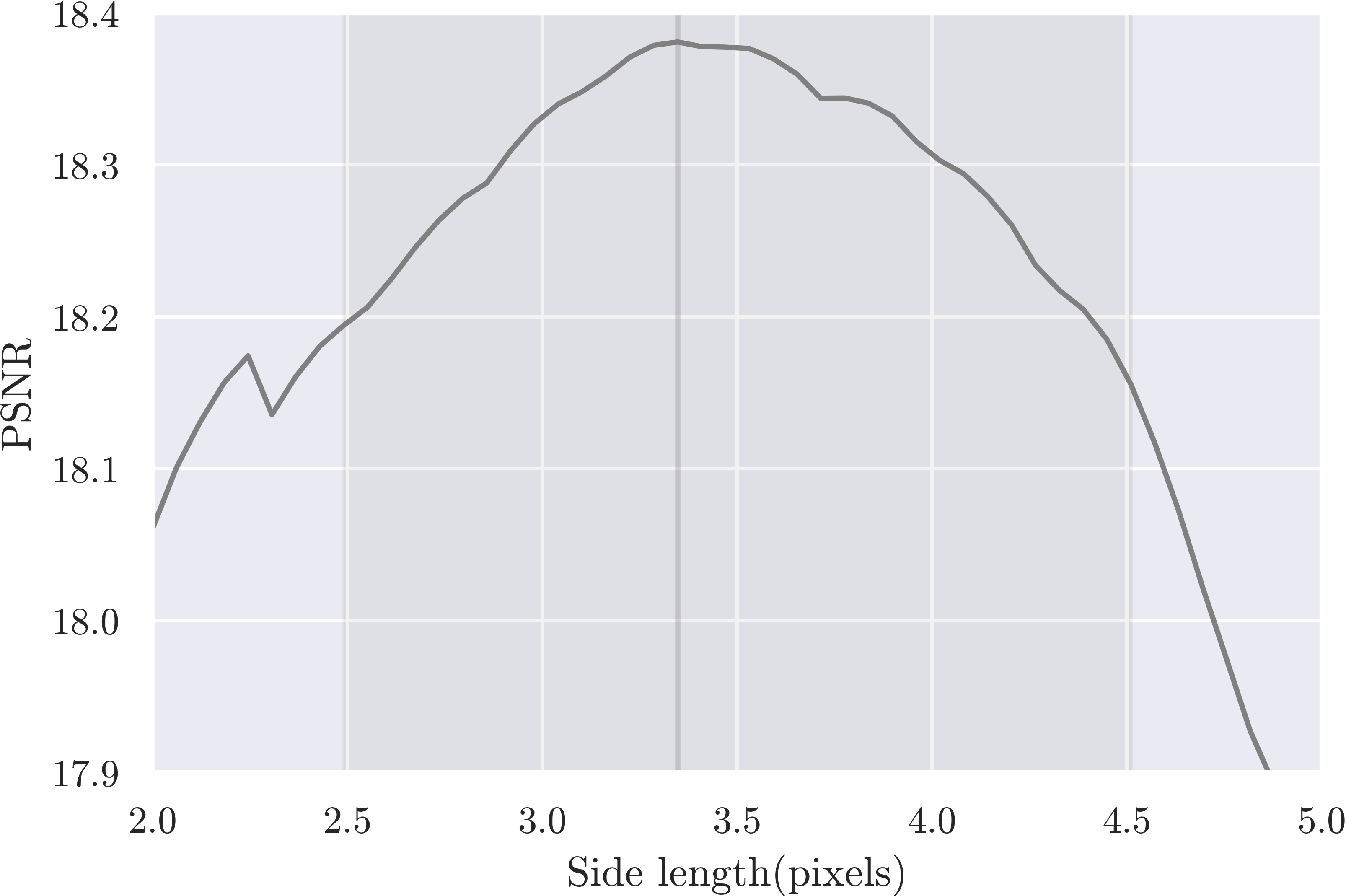}
        \caption{Square probe optimisation curve}
        \label{fig:probe_square}
    \end{subfigure}%
    \begin{subfigure}{0.5\textwidth}
        \includegraphics[width=\columnwidth,interpolate=false]{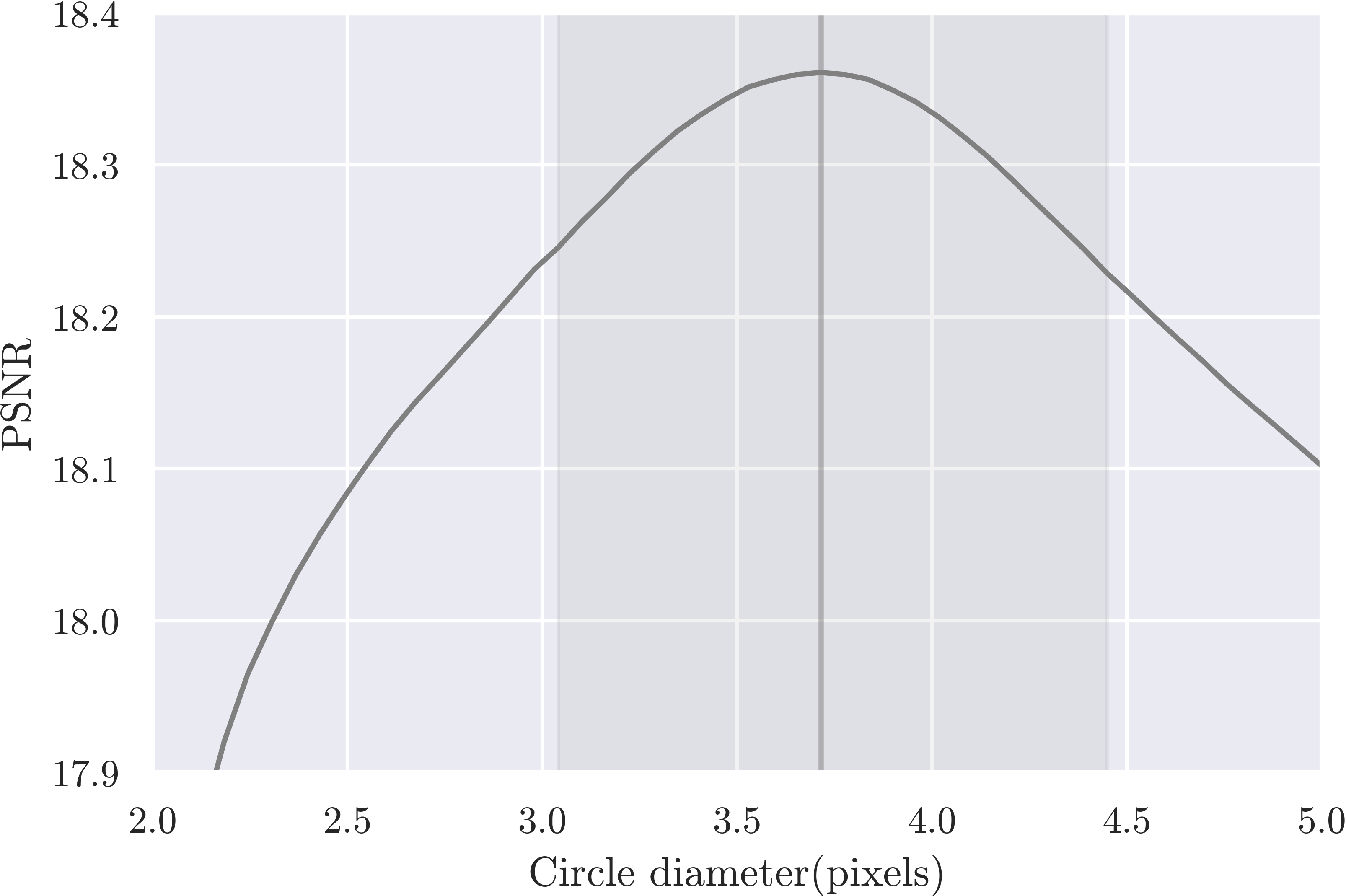}
        \caption{Circular probe optimisation curve}
        \label{fig:probe_circle}
    \end{subfigure}  
    \caption{The PSNR results of different probe sizes with a dark gray region indicating the 89\% HDI using the maximum average PSNR value as the reference. Figure~\ref{fig:probe_square} gives the optimisation curve for a square probe with the maximum average PSNR value of 18.38dB at 3.34 pixel side length.Figure~\ref{fig:probe_circle} gives the optimisation curve for a circular probe with the maximum average PSNR value of 18.36dB at 3.71 pixel diameter.}
    \label{fig:probes}
\end{figure}

The probe size of a scanning detector is not limited by the grid size as in the case of a staring array. One can select an optimal size for a square and circular probe by following a greedy optimisation approach. The approach scans from small to large probe size using the same sampling grid and compressed sensing to measure the average PSNR of the data set for each probe size (see Section~\ref{sec:method} for more information.). The results are shown in Figure~\ref{fig:probes} where the optimal probe size for the circular probe, a diameter of 3.71~pixels, and a side of 3.34~pixels for the square probe. The sampling approach, relative probe size, and regularisation algorithm for the circular probe is called the \textit{CS-Circle}, and for the square probe the \textit{CS-Square}.

\FloatBarrier


\subsection{Multi-Level Sampling Setup}
\label{sec:multi}
The extended theory of compressed sensing introduces multi-level sampling. In certain transformations, such as the Wavelet domain, levels can be interpreted as single coefficients within the transform or as a range of coefficients. Each level is then sampled from a uniform random distribution with a pre-determined number of samples. There is currently no guidelines in choosing the level ranges but it is suggested that samples are to be taken per level should it be distributed by a power law~\cite{Adcock2017}.

In the spatial domain, a single coefficient directly translates to a pixel value. Working with probes covering multiple pixels, the probe size can be interpreted as a level with the probe distributed uniformly across the output image. From this perspective, a range of levels will consist of a range of probe sizes. This interpretation also makes it possible to implement in hardware by changing the field of view of a detector.

In Section~\ref{sec:optimal}, a greedy approach was followed to find the single level providing optimal results. Using a greedy approach to find the optimal set of levels and adequate power law distribution in a large search space is not viable. We devise a model to be used to explore the parameter space and expected PSNR distribution of the multi-level sampling scheme.

The model uses five parameters. Four indicate a radius size; and one manages the coefficient of the power law which determines the number of samples per radius size. Let $r_i\in[0..31] \in\mathbb{R}$ be an arbitrary radius ranging from 0 to 31 pixels, using four different radiusses $i\in[0..3]\equiv R \in \mathbb{N}$ and requiring a total of $m\in\mathbb{N}$ samples. The number of samples $m_i\in\mathbb{R}$ per radius $r_i$ is given by
\begin{equation}
m_i = \left \lfloor{ c \cdot e^{-\alpha\left(\frac{i}{\|R\|}\right) } \cdot m }\right \rfloor
\end{equation}
where $c$ is a constant such that $\sum_i{e^{-\alpha\left(\frac{i}{\|R\|}\right) }} = 1$ and $m \leq \sum_i{m_i}$. The model parameter distributions are determined using the Markov Chain Monte Carlo (MCMC) algorithm called the ``t-walk"~\cite{Christen2010} with the PSNR value as the likelihood measure.

\section{Assessment methods and Protocols}
\label{sec:method}
The assessment of the various algorithms is performed by means of a data set, a quality measure and hypothesis testing to compare algorithms. A ``dead leaves" data set is generated with a power law distribution parameter set to 3 for the simulation and a small size cut-off on the circle radius at 0.7 pixels. This allows for a scale invariant approximation of natural scenes~\cite{Gousseau2008}. The data set contains 200 64x64 images. The quality measure for assessing image reconstruction accuracy is based on the PSNR measure. (See Appendix~\ref{app:psnr}). Images are converted to 16~bit, before measuring the PSNR, placing the maximum possible PSNR value at $96.33$~dB.

The algorithm comparison is performed by using two group comparisons using Bayesian estimation as presented by Kruschke~\cite{Kruschke2013}. A Student's t-distribution for model approximation and uniform priors are used (See Appendix~\ref{app:student}). The Highest Density Interval (HDI) of $89\%$ measured on the effect size is used to determine statistical significance~\cite{mcelreath2020statistical}. The relative performance of two algorithms are evaluated by approximating the posterior distribution of the effect size via a MCMC algorithm. Typically, negative effect size values support the first algorithm and positive effect size value supports the second algorithm. However, from a Bayesian perspective the algorithm is assumed to perform better if the 89\% HDI of the effect size are strictly positive or negative. 

\subsection{Grid-based Sampling}
Two evaluations are required. The first is to evaluate whether compressed sensing can be used as an interpolation technique and the second is to determine whether improved image formation performance can be achieved by optimising the probe size while sampling within a grid.

All three algorithms will be compared against each other as well as some well known interpolation techniques; namely nearest neighbour, linear, cubic and Lanczos interpolation. The interpolation techniques will upscale a 28x28 sized image which was decimated through an averaging process simulating staring array sampling. The seven algorithms will be compared using Bayesian estimation and the HDI of the effect size to determine whether any method is statistically better than another.

\subsection{Multi-level Sampling}
To investigate multi-level sampling, the PSNR distribution of the proposed model is compared to the PSNR distribution of single-level sampling. The model for single-level sampling contains a single parameter, indicating the probe radius and also uses the PSNR measure as a likelihood indicator. The two PSNR distributions will be compared to determine whether multi-level sampling does improve over single-level sampling in the spatial domain.

\section{Results}
\label{sec:results}
\subsection{Grid-based Sampling}
\begin{figure}[ht]
    \centering
    \begin{subfigure}{0.3\textwidth}
        \includegraphics[width=\columnwidth,interpolate=false]{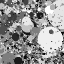}
        \caption{ Original \\ {\color{white} .} }
    \end{subfigure}\quad%
    \begin{subfigure}{0.3\textwidth}
        \includegraphics[width=\columnwidth,interpolate=false]{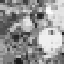}
        \caption{ Nearest \\ 16.52 [16.51]}
    \end{subfigure}\quad%
    \begin{subfigure}{0.3\textwidth}
        \includegraphics[width=\columnwidth,interpolate=false]{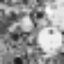}
        \caption{ Linear \\ 17.40 [17.38]}
    \end{subfigure}
    \begin{subfigure}{0.3\textwidth}
        \includegraphics[width=\columnwidth,interpolate=false]{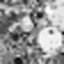}
        \caption{ Cubic \\ 17.95 [17.94]}
    \end{subfigure}\quad%
    \begin{subfigure}{0.3\textwidth}
        \includegraphics[width=\columnwidth,interpolate=false]{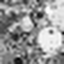}
        \caption{ Lanczos \\ 18.01 [17.99]}
    \end{subfigure}\quad%
    \begin{subfigure}{0.3\textwidth}
        \includegraphics[width=\columnwidth,interpolate=false]{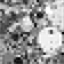}
        \caption{ CS-FPA \\ 18.18 [18.19]}
    \end{subfigure}
    \begin{subfigure}{0.3\textwidth}
        \includegraphics[width=\columnwidth,interpolate=false]{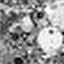}
        \caption{ CS-Circle \\ 18.36 [17.30]}
    \end{subfigure}\quad%
    \begin{subfigure}{0.3\textwidth}
        \includegraphics[width=\columnwidth,interpolate=false]{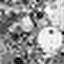}
        \caption{ CS-Square \\ 18.38 [18.35]}
    \end{subfigure}    
    \caption{Example images for each algorithm with the average PSNR value indicated below and the specific image PSNR shown in square brackets.}
    \label{fig:interpolation}
\end{figure}

An example image for each interpolation method and compressed sensing method is shown in Figure~\ref{fig:interpolation}; with the final results shown in Table~\ref{tab:results}.
\FloatBarrier

\begin{table}[ht!]
    \caption{The table provides the results for grid based sampling and interpolation techniques. The bold values give the average PSNR for each algorithm while all other values are the minimum HDI value calculated as row algorithm versus column algorithm. A negative HDI value indicates that the row algorithm is not statistically better than the column algorithm.}
    \label{tab:results}
    \begin{adjustbox}{width=\textwidth}
    \begin{tikzpicture}
        \matrix(dict)[matrix of nodes,
            nodes={align=center},
            row 1/.style={nodes={align=left,rotate=75,anchor=west,transform shape}},
            column 1/.style={nodes={text width=2cm,align=right}}
        ]{
            .         & Nearest & Linear & Cubic & Lanczos & CS-FPA & CS-Circle & CS-Square \\
            CS-Square &  2.71 & 1.34 & 0.50 & 0.38 & 0.13 &|[fill=gray!30]| -0.11 & \textbf{18.38(1.06)}\\
            CS-Circle & 2.64 & 1.30 & 0.41 & 0.30 &  0.05 & \textbf{18.32(1.07)}\\
            CS-FPA    & 2.39 & 1.03 & 0.16 & 0.06 & \textbf{18.18(1.08)}\\
            Lanczos    & 2.14 & 0.77 &|[fill=gray!30]| -0.08 & \textbf{18.01(1.08)}\\
            Cubic     & 2.04 & 0.66 & \textbf{17.94(1.08)}\\
            Linear    & 1.22 & \textbf{17.40(1.06)}\\
            Nearest   & \textbf{16.52(1.04)}\\
        };
        \draw(dict-2-1.north west)--(dict-2-8.north east);
        \draw(dict-8-1.south east)--([yshift=20pt]dict-2-1.north east);
        \draw[gray!50](dict-3-1.north west)--(dict-2-8.south east);
        \draw[gray!50](dict-4-1.north west)--(dict-3-7.south east);
        \draw[gray!50](dict-5-1.north west)--(dict-4-6.south east);
        \draw[gray!50](dict-6-1.north west)--(dict-5-5.south east);
        \draw[gray!50](dict-7-1.north west)--(dict-6-4.south east);
        \draw[gray!50](dict-8-1.north west)--(dict-7-3.south east);
    \end{tikzpicture}
    \end{adjustbox}
\end{table}

The numerical experimentation has shown that compressed sensing as an interpolation method~(CS-FPA) outperforms Lanczos interpolation as well as cubic interpolation. It is interesting to note that Lanczos interpolation does not outperform cubic interpolation in this case. Furthermore, the compressed sensing algorithms with optimised probes (CS-Circle and CS-Square) outperform all image formation algorithms but do not differ significantly from one another.

\subsection{Multi-level Sampling}
\begin{figure}[ht!]
    \centering
    \begin{subfigure}{0.9\textwidth}
        \includegraphics[width=\columnwidth,interpolate=false]{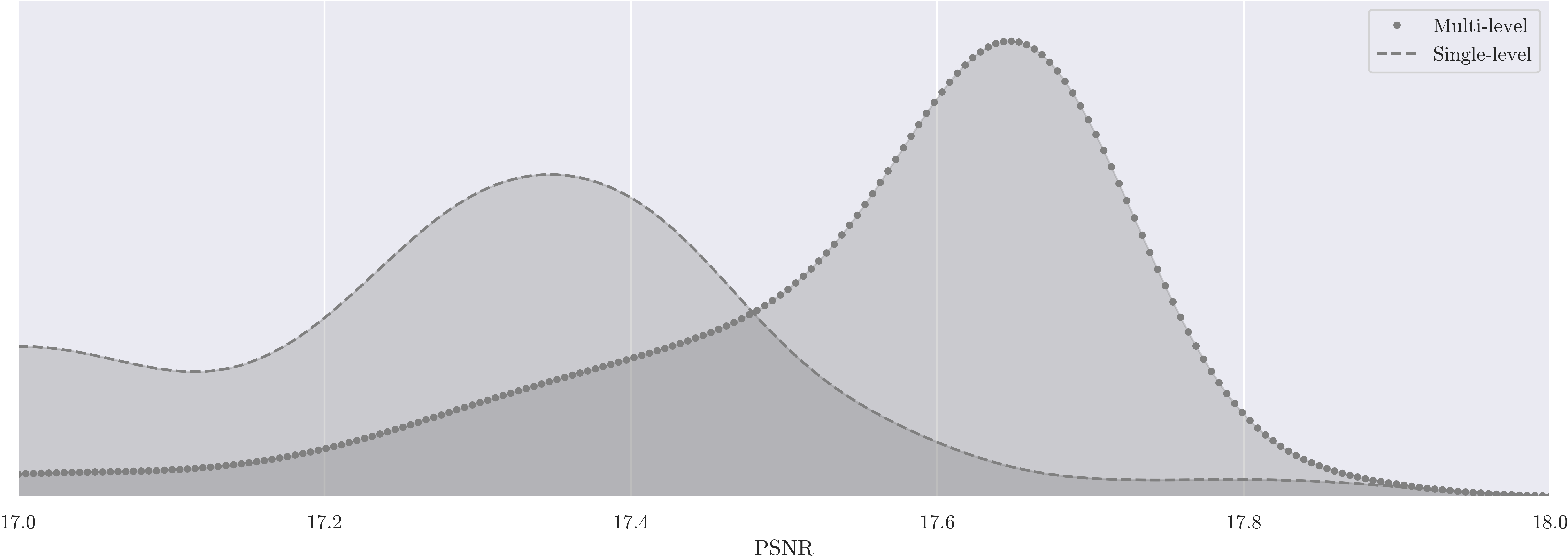}
        \caption{}
    \end{subfigure}%
    \caption{The PSNR distributions for multi-level and single-level sampling. The average PSNR value for multi-level sampling is 17.54dB and 17.27dB for single-level sampling.}
    \label{fig:distribution}
\end{figure}

\begin{figure}[ht!]
    \centering
    \begin{subfigure}{0.85\textwidth}
        \includegraphics[width=\columnwidth,interpolate=false]{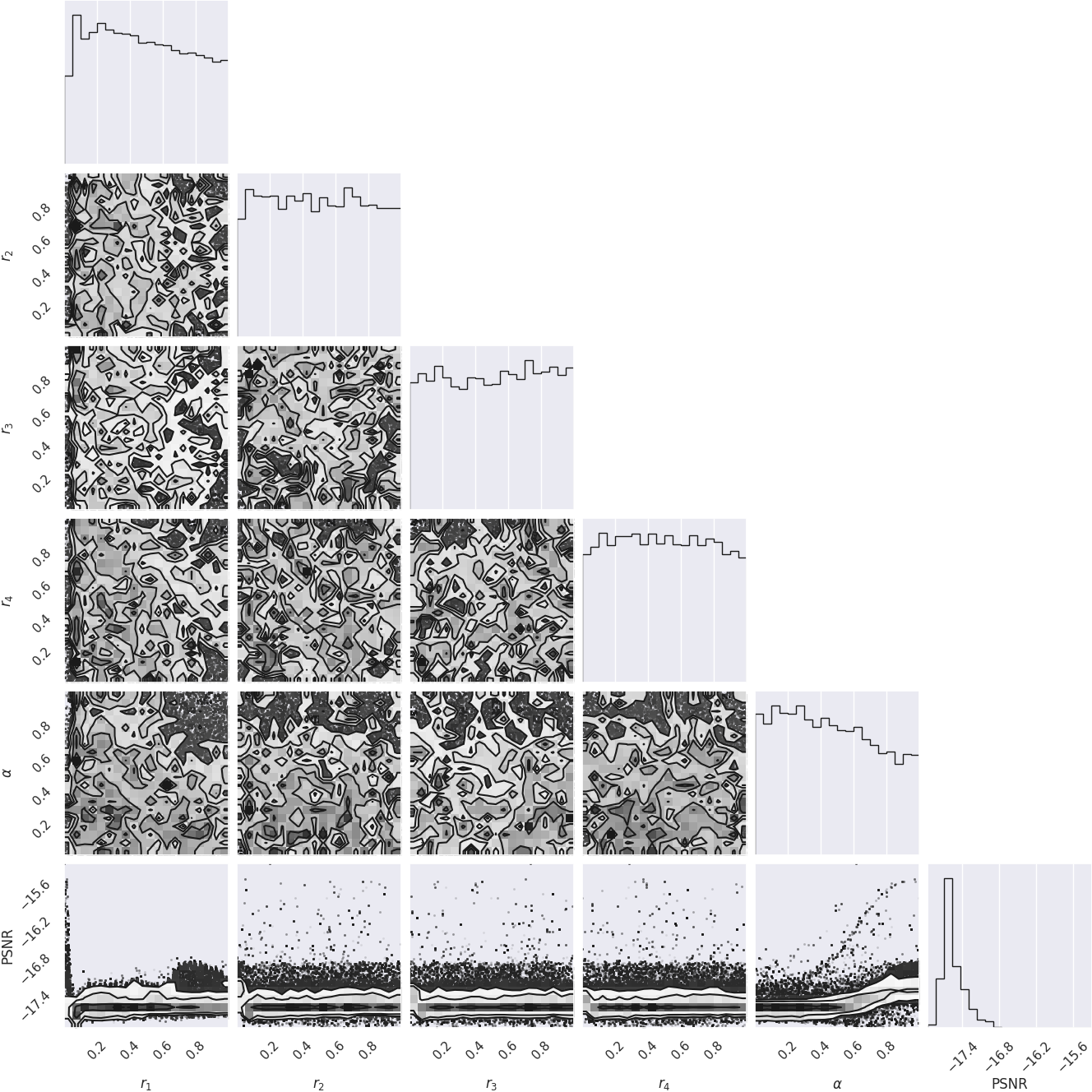}
        \caption{The parameter distributions for the multi-level model.}
        \label{fig:para6}
    \end{subfigure}
    \begin{subfigure}{\textwidth}
        \centering
        \includegraphics[width=0.3\columnwidth,interpolate=false]{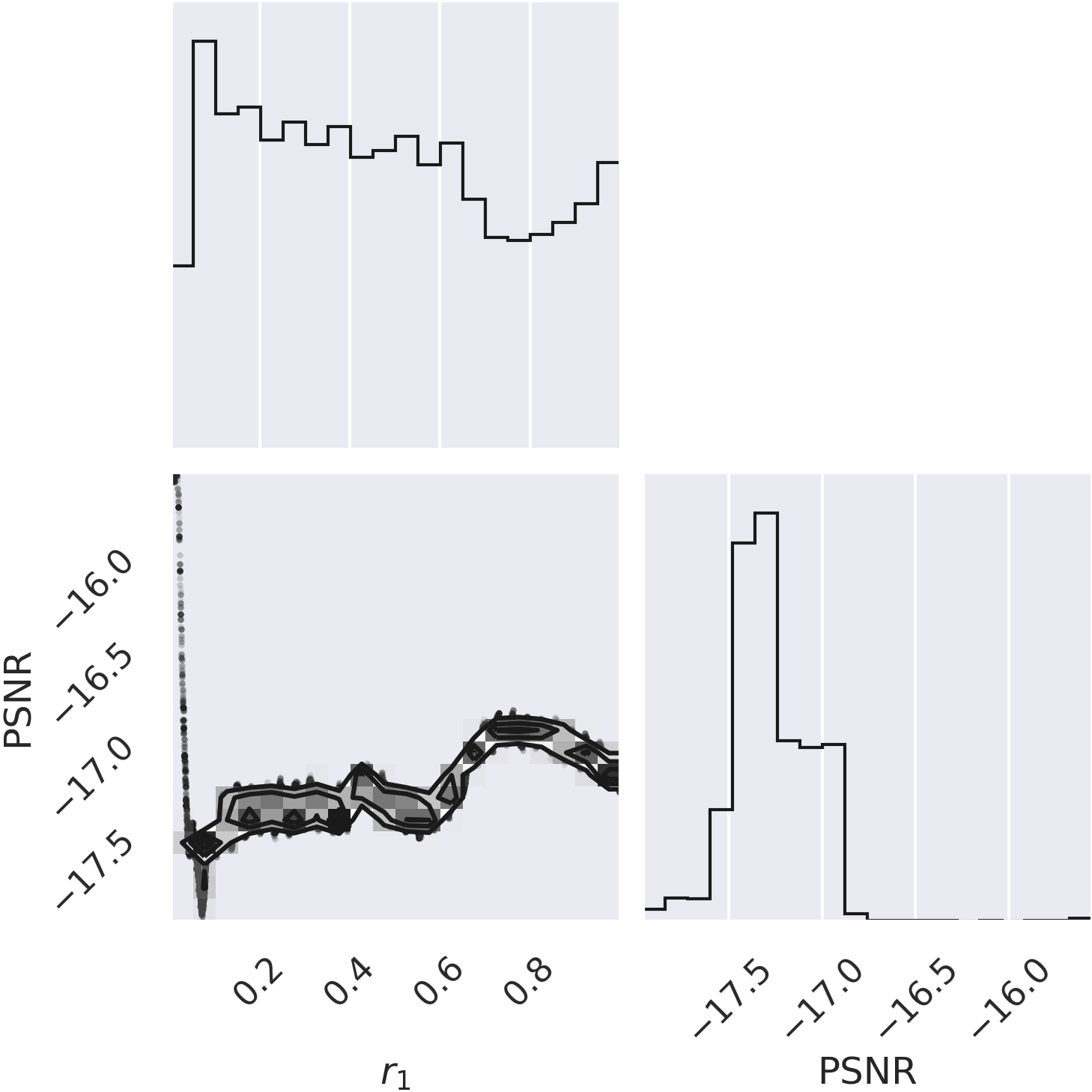}
        \caption{The parameter distributions for the single-level model.}
        \label{fig:para7}
    \end{subfigure}  
    \caption{The parameter co-variance matrices for the MCMC chains produced by the ''t-walk".}
    \label{fig:parameters}
\end{figure}

The PSNR distributions generated through the MCMC for single-level and multi-level sampling is shown in Figure~\ref{fig:distribution} while the  parameter distributions are shown in Figure~\ref{fig:parameters}.

The parameter distributions shown in Figure~\ref{fig:para7} show that only a small range of radii produce high PSNR values and, on average, all other parameters perform much worse. The parameter distributions in Figure~\ref{fig:para7} for $r_1$ is tending to the same parameter value found in the single-level sampling while the parameter $\alpha$ tends to be skewed to a lower value. Lower values of $\alpha$ create a flattened power law distribution indicating that values are more uniformly distributed over the four radius parameters. There is no correlation identified which will produce high PSNR values as in the single-level case. This might be due to the MCMC not finding adequate steps and might require custom adjustments.

\FloatBarrier

\section{Discussion and Future Work}
\label{sec:discuss}
It has been shown that single pixel scanning systems have the potential to outperform staring array based imaging systems in regards to image quality by adequately optimising the detectors and sampling patterns. It was first shown that compressed sensing can be used as an interpolation algorithm that outperforms traditional interpolation algorithms such as Lanczos and cubic interpolation. This was then extended to use a sample grid equivalent to a staring array but optimising the probe size used in the sampling. This further improved the image quality beyond what a staring array could produce for the same amount of samples. Finally it was shown that multi-level sampling improve its PSNR distribution compared to single-level sampling.

A 28x28 staring array was used and up-scaled to a 64x64 image by using an interpolation algorithm. The compressed sensing based interpolation algorithm, CS-FPA, was developed by modelling each detector in the staring array as a separate sample, with each sample corresponding to a square with sides equal to 2.28 pixels related to the output image. The chosen image reconstruction algorithm, FDRI, allows for generating an interpolation matrix usable on any image to upscale from 28x28 to 64x64. An interpolation matrix must be calculated for every set of up-scaling.  The CS-FPA outperformed Lanczos interpolation by producing an average PSNR value of 18.18~dB compared to 18.01~dB. It was then fathomed that if a single detector scanning system would sample at the same positions as the staring array, taking the centre of each element of the staring array as a position, it is possible to adjust the size of the sampling probe or detector field of view too possibly improve image quality. Not only can the probe size be changed but the probe shape can be selected as well. In the paper the probe shape was limited to a square and circular probe with algorithms called CS-Square and CS-Circle respectively. The optimal probe size was determined by selecting the probe size producing the highest average PSNR value. CS-Square with a probe size of 3.34~pixels produced a PSNR value of 18.38~dB while CS-Circle with a diameter of 3.71~pixels produced a PSNR value of 18.36dB; both outperformed CS-FPA and Lanczos.

It must be noted that CS-Circle and CS-Square is not interpolation algorithms as they are dependant on the physical sampling processes used. To summarise, when given a choice to take a set amount of samples with either a staring array or a single detector scanning system, and the detector field of view in the scanning system can be adjusted, the results presented here indicated that the scanning system will outperform the staring array.

Compressed sensing theory indicates that multi-level sampling can further improve signal reconstruction quality. Too investigate this, a multi-level sampling approach was developed and the PSNR distribution was compared to a single-level sampling approach. Only the PSNR distributions was compared due to the high parameter space present in multi-level sampling; there is currently no guarantee that optimal parameters will be found. The average PSNR distribution for multi-level sampling was improved compared to single-level sampling, 17.54~dB versus 17.17~dB respectively. It can be concluded that using a multi-level sampling model will improve the PSNR distribution compared to a single-level sampling model.

Spatial domain sampling can further be developed by investigating possible different probe forms and improved real-time regularisation algorithms. Multi-level sampling has shown potential and research in optimisation approaches for finding optimal sampling structures or identifying optimal probe ranges and corresponding number of samples can be investigated.

\appendix    

\section{Equations}
\subsection{Peak-Signal to Noise Ratio Equations}
\label{app:psnr}
Given an image $I$ of height $H$ and width $W$ such that $I(m, n) \mapsto i_{m,n} \in {0, \ldots,2^{16} - 1}$ with $|m| \leq H-1$, $|n| \leq W-1$ and $m,n \in \mathbb{Z}$. The Mean Square Error between a reference image $I_{R}$ and a test image $I_{T}$ is given be
\begin{equation*}
MSE(I_{T}, I_{R}) = \frac{1}{HW} 
\sum_{m=0}^{H-1}\sum_{n=0}^{W-1} \left[ I_{T}(m,n) - I_{R}(m,n)\right]^2
\end{equation*}
The Peak Signal-to-Noise Ratio~(PSNR) in decibel~(dB) is defined as
\begin{equation}
    \label{eqn:PSNR}
    PSNR(I_{T}, I_{R}) = 
    \begin{cases}
    20\cdot \log_{10}\left( \frac{2^{B} - 1}{\sqrt{MSE(I_{T}, I_{R})}}\right) &  MSE(I_{T}, I_{R}) \ge 1\\
    20\cdot \log_{10}\left( 2^{B} - 1 \right) & Otherwise
    \end{cases}
\end{equation}

\subsection{Student's t-distribution}
\label{app:student}
The Student's t-distribution is given by
\begin{equation*}
    \text{StudentT}(y|\nu,\mu,\sigma) = \frac{\Gamma\left((\nu + 1)/2\right)}      {\Gamma(\nu/2)} \ \frac{1}{\sqrt{\nu \pi} \ \sigma} \ \left( 1 + \frac{1}{\nu} \left(\frac{y - \mu}{\sigma}\right)^2 \right)^{-(\nu + 1)/2} \!
\end{equation*}
with $\nu$ as the normality parameter, $\mu$ as the mean, $\sigma$ as standard deviation and $y$ as the data point.

The effect size is calculated by:
\begin{equation*}
    \text{Effect Size} = \frac{\left( \mu_1 - \mu_2 \right) }{\sqrt{\frac{1}{2}\left( \sigma_1^2 + \sigma_2^2\right)}}
\end{equation*}

 

\bibliography{references}
\bibliographystyle{spiebib}

\end{document}